\documentclass[twocolumn,showpacs,preprintnumbers,amsmath,amssymb,floatfix]{revtex4}

\usepackage[dvipdfmx]{graphicx}
\usepackage{bm}
\usepackage{color}
\usepackage{amsmath}
\usepackage{amsfonts}
\usepackage{amssymb}
\usepackage{ulem}

\usepackage[colorlinks=true,bookmarksnumbered=true,citecolor=blue,linkcolor=blue,urlcolor=blue,bookmarks=false]{hyperref}

\begin{document}

\preprint{APS/123-QED}

\title{Structures of magnetic excitations in the spin-1/2 kagome-lattice antiferromagnets Cs$_2$Cu$_3$SnF$_{12}$ and Rb$_2$Cu$_3$SnF$_{12}$}

\author{Mutsuki\,\,Saito$^1$}
\email{saito.m.bc@m.titech.ac.jp}
\author{Ryunosuke\,\,Takagishi$^1$}
\author{Nobuyuki\,\,Kurita$^1$}
\author{Masari\,\,Watanabe$^1$} 
\author{Hidekazu\,\,Tanaka$^1$} 
\email{tanaka.h.ag@m.titech.ac.jp}
\author{Ryuji\,\,Nomura$^2$} 
\author{Yoshiyuki\,\,Fukumoto$^3$}
\author{Kazuhiko\,\,Ikeuchi$^4$}
\author{Ryoichi\,\,Kajimoto$^5$}
\affiliation{
$^1$Department of Physics, Tokyo Institute of Technology, Meguro-ku, Tokyo 152-8551, Japan\\
$^2$Graduate School of Engineering, Hokkaido University, Sapporo, Hokkaido 060-8628, Japan\\
$^3$Department of Physics, Faculty of Science and Technology, Tokyo University of Science, Noda, Chiba 278-8510, Japan\\
$^4$Comprehensive Research Organization for Science and Society (CROSS), Tokai, Ibaraki 319-1106, Japan\\
$^5$Materials and Life Science Division, J-PARC Center, Japan Atomic Energy Agency, Tokai, Ibaraki 319-1195, Japan
}

\date{\today}

\begin{abstract}
Cs$_2$Cu$_3$SnF$_{12}$ is a spin-1/2 antiferromagnet on a nearly uniform kagome lattice with an exchange interaction of $J\,{=}\,20.7$ meV. This compound undergoes magnetic ordering at $T_{\mathrm{N}}\,{=}\,20.2$ K with the $\bm{q}\,{=}\,0$ structure and the positive chirality, which is mainly caused by the large Dzyaloshinsky--Moriya interaction. Rb$_2$Cu$_3$SnF$_{12}$ is a spin-1/2 antiferromagnet on a modified kagome lattice with a $2a{\times}2a$ enlarged chemical unit cell at room temperature resulting in four kinds of nearest-neighbor exchange interaction, the average of which is $J_{\mathrm{avg}}\,{=}\,15.6$ meV. Its ground state is a pinwheel valence bond solid (VBS) with an excitation gap. Here, we show the structures of magnetic excitations in Cs$_2$Cu$_3$SnF$_{12}$ and Rb$_2$Cu$_3$SnF$_{12}$ investigated by inelastic neutron scattering in wide energy and momentum ranges. For Cs$_2$Cu$_3$SnF$_{12}$, four single-magnon excitation modes were observed. Low-energy three modes are assigned to be transverse modes and the high-energy fourth mode is suggested to be an amplitude mode. It was confirmed that the energy of single-magnon excitations arising from the $\Gamma'$ point in the extended Brillouin zones is largely renormalized downwards. It was found that the broad excitation continuum without a marked structure spreads in a wide energy range from $0.15J$ to approximately $2.5J$ in contrast to the clearly structured excitation continuum observed in the spin-1/2 triangular-lattice Heisenberg antiferromagnet. These findings, as well as the results of the recent theories based on fermionic approach of spinon excitations from the spin liquid ground state, strongly suggest spinon excitations as elementary excitations in Cs$_2$Cu$_3$SnF$_{12}$. In Rb$_2$Cu$_3$SnF$_{12}$, singlet--triplet excitations from the pinwheel VBS state and their ghost modes caused by the enlargement of the chemical unit cell were clearly confirmed. It was found that the excitation continuum is structured in the low-energy region approximately below $J_{\mathrm{avg}}$ and the almost structureless high-energy excitation continuum extends to approximately $2.6J_{\mathrm{avg}}$. The characteristics of the high-energy excitation continuum are common to both Cs$_2$Cu$_3$SnF$_{12}$ and Rb$_2$Cu$_3$SnF$_{12}$, irrespective of their ground states. 
The experimental results strongly suggest that the spin liquid component remains in the ground state as quantum fluctuations in Cs$_2$Cu$_3$SnF$_{12}$ and Rb$_2$Cu$_3$SnF$_{12}$.

\end{abstract}

\maketitle


\section{Introduction}\label{Intro}
Linear spin wave theory (LSWT) is a basis for understanding magnetic excitations from the magnetically ordered state~\cite{Holstein,Anderson,Kubo}. Hence, LSWT has been widely applied to analyze magnetic excitations in magnetic materials. The magnetic excitations in most three-dimensional (3D) magnets, which are describable using the classical spin model, can be well understood in terms of LSWT. The characteristic elementary excitation of LSWT is a spin-1 excitation, magnon. Over the last two decades, the magnetic excitations in frustrated quantum magnets, such as spin-1/2 triangular-lattice Heisenberg antiferromagnets (TLHAFs) and kagome-lattice Heisenberg antiferromagnets (KLHAFs), have been drawing considerable attention from the viewpoint of fractionalized spin excitations~\cite{Zheng,Mezio,Ghioldi,Ghioldi2,Ferrari,Zhang,Punk,Zhang2,Ferrari2}. A typical fractionalized spin excitation is a spin-1/2 excitation, spinon, which has been established as the elementary excitation of spin-1/2 antiferromagnetic Heisenberg chain~\cite{dCP,Faddeev}.

Magnetic excitations of $S\,{=}\,1/2$ TLHAFs have been actively investigated theoretically using various approaches~\cite{Ghioldi,Ghioldi2,Ferrari,Zhang,Starykh,Zheng,Chernyshev,Mezio,Mourigal}. The theories demonstrated that although the dispersion relation of low-energy single-magnon excitations near the K point can be described by LSWT, the excitation energy is significantly renormalized downward by quantum fluctuations in a large area of the Brillouin zone (BZ)~\cite{Starykh,Zheng,Chernyshev,Mezio,Mourigal}, and that the dispersion curve shows a rotonlike minimum at the M point~\cite{Zheng,Mezio,Verresen}. These theoretical predictions were confirmed by inelastic neutron scattering (INS) experiments performed on Ba$_3$CoSb$_2$O$_9$~\cite{Ma,Ito,Kamiya,Macdougal}, which closely approximates the ideal $S\,{=}\,1/2$ TLHAF~\cite{Shirata,Zhou,Susuki,Koutroulakis,Yamamoto}. The rotonlike minimum in the single dispersion curve was observed not only at the M point but also at the Y point, the middle point between $\Gamma$ and M points~\cite{Ito,Macdougal}. These dispersion anomalies were successfully reproduced by fermionic approaches based on resonating valence bond (RVB) theory or a closely related theory~\cite{Ferrari,Zhang}, in which an elementary spin excitation is the spinon and the single-magnon excitation is described by the bound state of spinons. 

Experiments on Ba$_3$CoSb$_2$O$_9$ revealed that the excitation spectrum has an intense structured continuum extending to the high-energy region, which is at least six times higher than the exchange interaction $J$~\cite{Ito,Macdougal}. The structured intense continuum cannot be explained in terms of two-magnon excitations, because the calculated intensity is much smaller than the observed intensity~\cite{Kamiya}. In the theory based on spinon excitations, single-magnon excitations and the excitation continuum are described by the bound state of spinons and the spinon continuum, respectively~\cite{Ghioldi,Ghioldi2,Ferrari,Zhang}. The intense excitation continuum can be described by the spinon theory, although there is still disagreement between theory and experiment about the structure of the excitation spectrum. Experimental and theoretical results suggest that the dominant elementary excitations in the $S\,{=}\,1/2$ TLHAF are spinon excitations. Such fractionalized spin excitations will be more likely to occur in the $S\,{=}\,1/2$ KLHAF than in the $S\,{=}\,1/2$ TLHAF, because the ground state of the $S\,{=}\,1/2$ KLHAF has been considered to be a quantum disordered state.

The $S\,{=}\,1/2$ KLHAF with the nearest-neighbor interaction has been extensively studied and is a research frontier in condensed matter physics even today. It was theoretically demonstrated that for large spins, quantum fluctuation stabilizes the $\sqrt{3}\times\sqrt{3}$ structure~\cite{Chubukov,Reimers}, whereas for $S\,{=}\,1/2$, the synergistic effect of strong frustration and quantum fluctuation leads to the quantum disordered ground state \cite{Zeng1,Sachdev,Chalker,Elstner,Nakamura,Zeng2,Lecheminant,Waldtmann,Mila,Mambrini,Hida,Syromyatnikov,Jiang,Nishimoto,Capponi,Schnack}. However, the nature of the ground state for the spin-1/2 KLHAF is still under debate. Valence bond crystals (VBCs) described by a static array of singlet dimers \cite{Nikolic,Budnik,Singh1,Singh2,Yang,Hwang1} and quantum spin liquid based on the RVB theory are representative models of the ground state. Most of recent theories support the spin liquid ground state, which includes the gapless $U(1)$ Dirac spin liquid state~\cite{Hastings,Ran,Hermele,Iqbal,Liao,He,Zhu,Jiang2} and gapped $Z_2$ spin liquid state~\cite{Jiang,Yan,Depenbrock,Jiang4,Mei}. 

\begin{figure*}[t]
\includegraphics[width=15cm, clip]{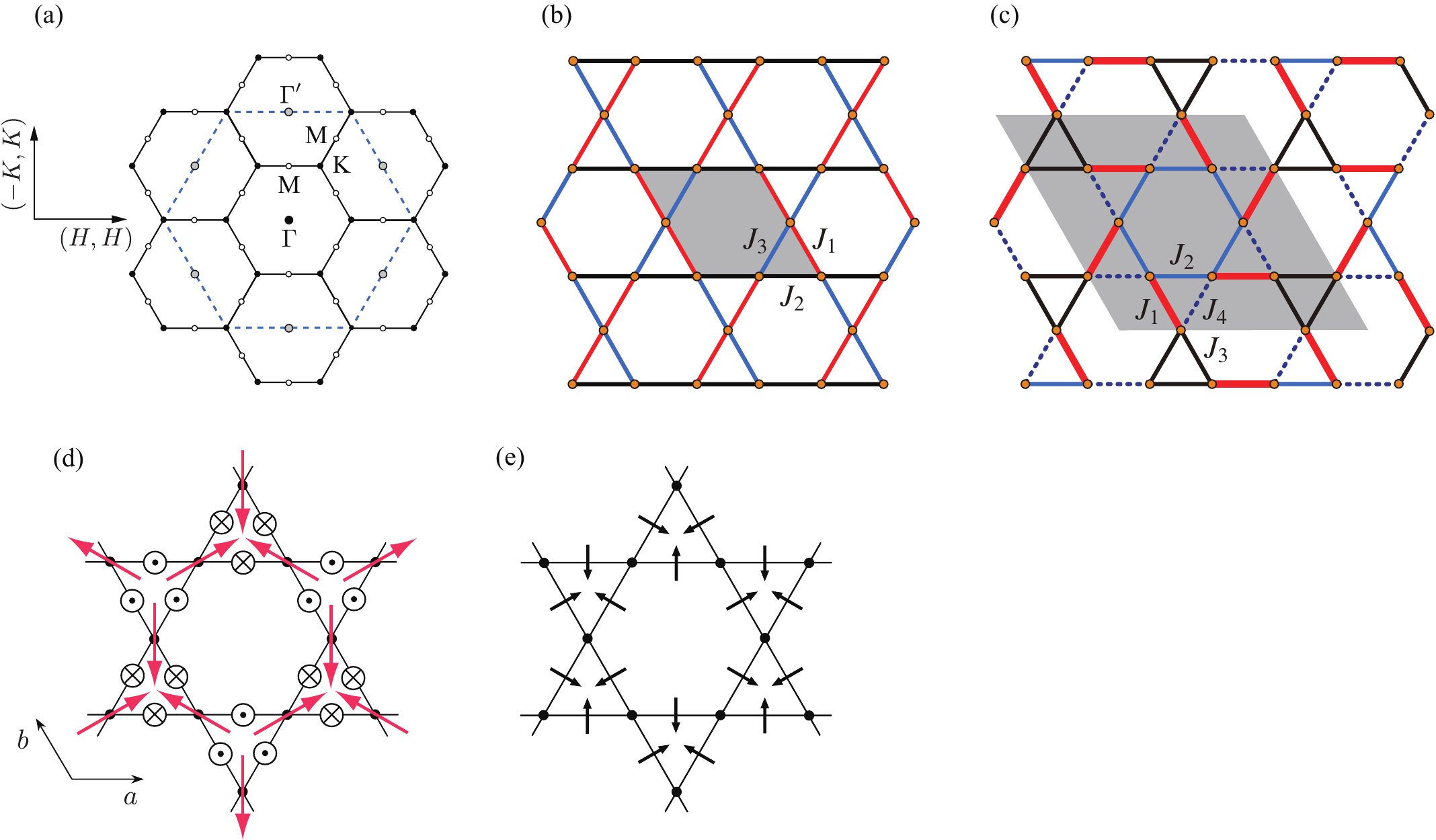}
\caption{(Color online) (a) 2D reciprocal lattice for the kagome lattice. Hexagons drawn with solid lines show the elementary BZs. An extended BZ is drawn with dashed lines. The reciprocal lattice points given by $(2m, 2n)$ with integers $m$ and $n$, and those given by $(2m+1, n)$ or $(m, 2n+1)$ are labeled as $\Gamma$ and $\Gamma'$, respectively. (b) Exchange network below $T_{\mathrm{t}}\,{=}\,185$ K in the kagome layer of Cs$_2$Cu$_3$SnF$_{12}$~\cite{Matan3}. There are three kinds of nearest-neighbor exchange interactions $J_1, J_2$, and $J_3$, which are expected to be almost the same from their bonding angles $\mathrm{Cu}^{2+}-\mathrm{O}^{2-}-\mathrm{Cu}^{2+}$. (c) Exchange network at room temperature in the kagome layer of Rb$_2$Cu$_{3}$SnF$_{12}$, in which pinwheels composed of exchange interactions $J_1$, $J_2$, and $J_4$ are connected by triangles of $J_3$ interactions~\cite{Morita,Matan}. Shaded areas in (b) and (c) are the unit cells for Cs$_2$Cu$_3$SnF$_{12}$ and Rb$_2$Cu$_3$SnF$_{12}$, respectively, at room temperature. (d) and (e) Arrangement of the $c$ axis component $D^\parallel$ and the $c$ plane component $D^\perp$, respectively, of the $\bm{D}$ vectors for the DM interaction at room temperature in Cs$_2$Cu$_3$SnF$_{12}$. Large arrows in (d) denote the $\bm{q}\,{=}\,0$ structure with positive chirality stabilized when $D^\parallel>0$.}
\label{fig:kagome_latt}
\end{figure*} 

Theoretical studies of magnetic excitations in $S\,{=}\,1/2$ KLHAF seem less extensive, because the ground state nature is unclear. Regarding experimental studies of $S\,{=}\,1/2$ KLHAFs, on the other hands, herbertsmithite ZnCu$_3$(OH)$_6$Cl$_2$ with the nearest-neighbor exchange interaction of $J\,{\simeq}\,17$ meV has been studied most extensively~\cite{Shores,Mendels,Helton,Bert,Lee,Imai,Olariu,Vries,Mendels2,Han2,Fu,Zorko,Khuntia}. The magnetic excitations in herbertsmithite were investigated via INS~\cite{Han,Han3}. The excitation spectrum is broad and almost featureless except that the scattering intensity is absent in the BZ centered at the $\Gamma$ points shown in Fig.~\ref{fig:kagome_latt}\,(a). The periodic absence of intensity around the $\Gamma$ points generally arises from the kagome geometry. The experimental results stimulated theoretical work on the magnetic excitations in $S\,{=}\,1/2$ KLHAF~\cite{Punk,Zhang2,Ferrari2,Prelovsek}. Recent theories based on the spin liquid ground state demonstrated that the strong lowest excitations composed of spinons occur at $\Gamma'$ points~\cite{Punk,Zhang2,Ferrari2}.  

Although herbertsmithite has been studied most extensively, this system has a problem that 15\% of Zn$^{2+}$ sites are substituted for Cu$^{2+}$ ions~\cite{Han2,Han3}, which leads to the exchange randomness in the kagome layer~\cite{herbertsmithite}.  
According to a theory, the exchange randomness smears an excitation spectrum making it broad~\cite{Shimokawa}. For a comprehensive study of the magnetic excitations in $S\,{=}\,1/2$ KLHAF, a model substance without exchange randomness is necessary. INS studies on other $S\,{=}\,1/2$ kagome-lattice antiferromagnets such as vesignieite~\cite{Boldrin} and kapellasite~\cite{Fak} have been conducted using powder samples. However, it seems difficult to obtain the detailed structure of the excitation spectrum from powder scattering data. 
 
Here, we show the structures of magnetic excitations in Cs$_2$Cu$_3$SnF$_{12}$ and Rb$_2$Cu$_3$SnF$_{12}$ investigated by INS in the wide energy and momentum ranges using single crystals. The advantages of these compounds are that large single crystals can be obtained and that the exchange randomness is absent. Cs$_2$Cu$_3$SnF$_{12}$ is magnetically described as an $S\,{=}\,1/2$ KLHAF with the nearest-neighbor exchange interaction of $J\,{=}\,20.7$ meV and the Dzyaloshinsky--Moriya (DM) interaction~\cite{Ono,Ono2}. At room temperature, this compound has a trigonal structure ($R\bar{3}m$) with a uniform kagome lattice of Cu$^{2+}$. Since the CuF$_6$ octahedra are tetragonally elongated approximately parallel to the $c$ axis owing to the Jahn--Teller effect, the hole orbitals $d(x^2-y^2)$ of Cu$^{2+}$ are spread in the kagome layer. This leads to a strong superexchange interaction in the kagome layer and a weak superexchange interaction between layers.  

Cs$_2$Cu$_3$SnF$_{12}$ undergoes a structural phase transition to a monoclinic structure ($P2_1/n$) at $T_{\mathrm{t}}\,{=}\,185$ K~\cite{Ono}, below which the kagome lattice is distorted~\cite{Ono2,Downie2,Matan3}. The temperature dependence of magnetic susceptibility exhibits a small kink anomaly at $T_{\mathrm{t}}$. There are three kinds of nearest-neighbor exchange interactions below $T_{\mathrm{t}}$, as shown in Fig.~\ref{fig:kagome_latt}\,(b)~\cite{Downie2,Matan3}. However, precise structural analysis revealed that the bonding angles $\mathrm{Cu}^{2+}-\mathrm{O}-\mathrm{Cu}^{2+}$ for these three kinds of exchange interactions are almost the same~\cite{Matan3}. This indicates that the kagome lattice below $T_{\mathrm{t}}$ can be regarded to be approximately uniform. The magnetic susceptibility of Cs$_2$Cu$_3$SnF$_{12}$ actually coincides with the theoretical magnetic susceptibility calculated by 24-site exact diagonalization~\cite{Misguich2} down to 30 K, which is just below the temperature $T_{\rm max}\,{\simeq}\,(1/6)J/k_{\rm B}\,{\simeq}\,40$\,K displaying a rounded maximum, as shown in Ref.~\cite{Ono}. 
 
Cs$_2$Cu$_3$SnF$_{12}$ undergoes a magnetic phase transition at $T_{\mathrm{N}}\,{=}\,20.0$ K to a $\bm{q}\,{=}\,0$ state with positive chirality owing to the relatively large DM interaction ($D^\parallel\simeq0.2J$), which conjugates with the ordered state~\cite{Ono,Ono2,Matan3,Cepas,Lee2}, and the weak interlayer exchange interaction $J'$, which is estimated to be $J'/J\,{\sim}\,1{\times}10^{-3}$ using a theory by Yasuda {\it et al}.~\cite{Yasuda}.

Ono \textit{et al}.~\cite{Ono2} reported on magnetic excitations in Cs$_2$Cu$_3$SnF$_{12}$ investigated by INS using a triple-axis spectrometer. They observed single-magnon excitations below 14 meV. Using LSWT, they analyzed the dispersion relations on the basis of a $2a\,{\times}\,2a$ enlarged chemical unit cell expected below $T_{\mathrm{t}}\,{=}\,185$\,K and found the large downward renormalization of excitation energy. However, the high-energy single-magnon excitations and excitation continuum in Cs$_2$Cu$_3$SnF$_{12}$ have not been examined. 

Rb$_2$Cu$_3$SnF$_{12}$ is also magnetically described as an $S\,{=}\,1/2$ KLHAF with a modified kagome lattice. At room temperature, this compound has a trigonal structure ($R\bar{3}$)~\cite{Morita}. The chemical unit cell is $2a\times2a$ enlarged in the $ab$ plane; thus, there are four kinds of the nearest-neighbor exchange interaction $J_\alpha$ ($\alpha\,{=}\,1$-$4$). As shown in Fig.~\ref{fig:kagome_latt}\,(c), exchange interactions $J_1$, $J_2$, and $J_4$ form pinwheels, which are linked by triangles of $J_3$ interactions~\cite{Morita,Matan}. Here, the exchange interaction $J_\alpha$ is labeled in decreasing order in magnitude. Rb$_2$Cu$_3$SnF$_{12}$ undergoes a structural phase transition at $T_{\mathrm{t}}\,{=}\,215$ K~\cite{Matan,Downie}. The low-temperature crystal structure has been considered to be a triclinic ($P\bar{1}$)~\cite{Downie}. However, the change in the exchange network at $T_{\mathrm{t}}$ should be small because the magnetic susceptibility exhibits no anomaly at $T_{\mathrm{t}}$~\cite{Morita}.

The magnetic ground state of Rb$_2$Cu$_3$SnF$_{12}$ is a pinwheel valence bond solid (VBS) with an excitation gap of $\Delta\,{=}\,2.4$ meV~\cite{Morita,Matan}. 
From the analysis of the dispersion curves obtained by INS experiment, the exchange constants were determined to be $J_1\,{=}\,18.6$\,meV, $J_2\,{=}\,0.95J_1$, $J_3\,{=}\,0.85J_1$, $J_4\,{=}\,0.55J_1$, and $D_\alpha^\parallel\,{=}\,0.18J_\alpha$~\cite{Matan}, which are consistent with those estimated from the temperature dependence of magnetic susceptibility~\cite{Morita}. The average of four exchange constants is $J_{\mathrm{avg}}\,{=}\,15.6$ meV. Although low-energy singlet--triplet excitations in Rb$_2$Cu$_3$SnF$_{12}$ have been investigated in detail, high-energy singlet--triplet excitations and the excitation continuum are still unclear. 

In this paper, we focus on high-energy single-magnon excitations and the excitation continuum, which reflect the characteristics of magnetic quasiparticles. For Cs$_2$Cu$_3$SnF$_{12}$, we observed four single-magnon excitation modes around the $\Gamma'$ point in the BZ. Low-energy three modes are transverse modes that can be seemingly understood in terms of LSWT. The high-energy fourth mode is suggested to be a longitudinal mode composed of bound spinons as has been discussed by the RVB theory of magnetic excitations in $S\,{=}\,1/2$ TLHAF~\cite{Zhang}. We confirmed that the energy of single-magnon excitations arising from the $\Gamma'$ point is largely renormalized downwards~\cite{Ono2}. We found that the high-energy single-magnon excitations near the zone boundary decay significantly. We also found that the excitation continuum without marked structure spreads from $0.15J$ to at least $2.5J$ in contrast to the structured excitation continuum observed in the spin-1/2 triangular-lattice Heisenberg-like antiferromagnet Ba$_3$CoSb$_2$O$_9$~\cite{Ma,Ito,Kamiya,Macdougal}. In addition, unexpected weak excitations arising from the M points were observed. We will discuss the possible origins of the weak excitations. For Rb$_2$Cu$_3$SnF$_{12}$, we confirmed singlet--triplet excitations and their ghost modes attributable to the enlargement of the chemical unit cell in the $ab$ plane. We found a structured low-energy excitation continuum, which appears to be caused by the two triplet excitations, and an almost structureless high-energy continuum extending to at least $2.6J_{\mathrm{avg}}$. These characteristics of the high-energy excitation continuum are common to Cs$_2$Cu$_3$SnF$_{12}$ and Rb$_2$Cu$_3$SnF$_{12}$, although their ground states and low-energy excitations are different; thus, these characteristics of high-energy excitation continuum are considered to be universal in the $S\,{=}\,1/2$ KLHAF. The results obtained for both systems are discussed referring to the theory based on spinon excitations from the spin liquid ground state~\cite{Punk,Zhang2,Ferrari2}.

\section{Experimental Details}\label{Exp}
$A_2$Cu$_3$SnF$_{12}$ with $A\,{=}\,\mathrm{Cs}$ and Rb single crystals were synthesized in accordance with the chemical reaction 2$A$F + 3CuF$_2$ + SnF$_4 \rightarrow$ $A_2$Cu$_3$SnF$_{12}$. $A$F, CuF$_2$, and SnF$_4$ were dehydrated by heating in vacuum at about 100$^\circ$C. First, the materials were packed into a Pt tube of 15 or 10 mm inner diameter and 100 mm length at a ratio of $3\,{:}\,3\,{:}\,2$. Both ends of the Pt tube were tightly folded with pliers and placed between Nichrome plates. Single crystals were grown from the melt. The temperature of the furnace was lowered from 850 to 750$^\circ$C for Cs$_2$Cu$_3$SnF$_{12}$ and 800 to 700$^\circ$C for Rb$_2$Cu$_3$SnF$_{12}$ over 100 h. After collecting the well-formed crystals, we repeated the same procedure using a Pt tube of 13 or 10 mm inner diameter and 100 mm length. Transparent light-green crystals were obtained. 

Magnetic excitations of Cs$_2$Cu$_3$SnF$_{12}$ and Rb$_2$Cu$_3$SnF$_{12}$ were measured in a wide momentum-energy range using the Fermi chopper spectrometer 4SEASONS~\cite{Kajimoto} installed in the Materials and Life Science Experimental Facility (MLF) at J-PARC, Japan. We used one single crystal in each INS experiment. The size of the single crystal is $4\times1\times0.5$ cm$^3$ for Cs$_2$Cu$_3$SnF$_{12}$ and $3.5\times0.7\times0.3$ cm$^3$ for Rb$_2$Cu$_3$SnF$_{12}$. Their mosaicities are approximately $2^{\circ}$ and $1.5^{\circ}$, respectively. The sample was mounted in a cryostat with its $(1,1,0)$ and $(0,0,1)$ directions in the horizontal plane. The sample was cooled to 5 K using a closed-circle helium refrigerator. First, we checked the excitation spectra of Cs$_2$Cu$_3$SnF$_{12}$ and Rb$_2$Cu$_3$SnF$_{12}$, setting the wave vector $\bm{k}_{\mathrm{i}}$ of an incident neutron parallel to the $(0,0,1)$ direction. Scattering data presented in this paper were collected by rotating the sample around the $(-1,1,0)$ direction with a set of incident neutron energies: 6.3, 8.5, 11.9, 17.9, 30.0, and 60.0 meV~\cite{Nakamura2}. All the data were analyzed using the software suite UTSUSEMI~\cite{Inamura}.

\begin{figure*}[t]
\includegraphics[width=17.5cm, clip]{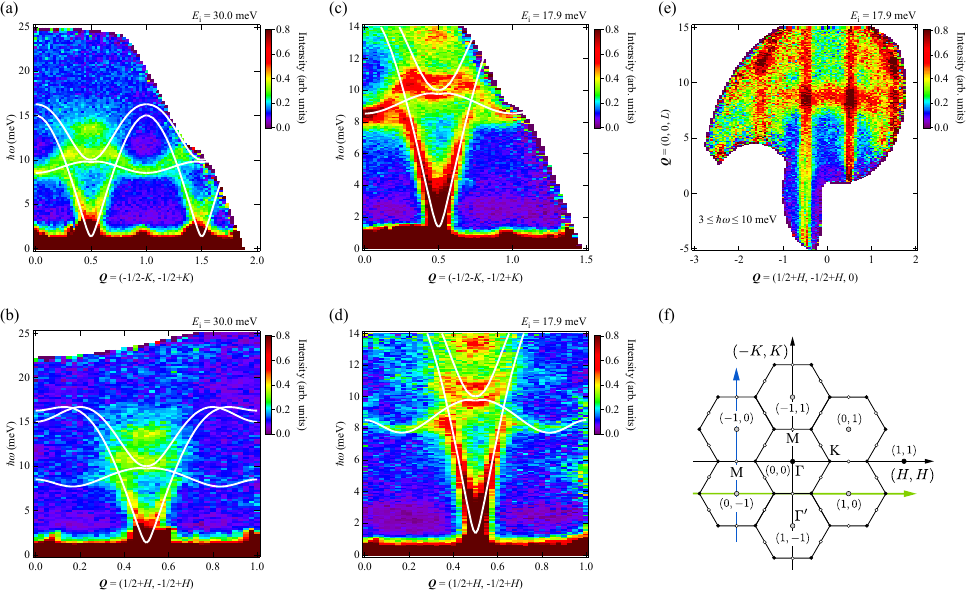}
\caption{(Color online) Excitation spectra of Cs$_2$Cu$_3$SnF$_{12}$ measured at $T\,{=}\,5$ K. (a)--(d) Energy--momentum maps of scattering intensity along two high-symmetry directions $\bm{Q}\,{=}\,(-\frac{1}{2}-K, -\frac{1}{2}+K)$ (blue line) (a), (c) and $(\frac{1}{2}+H, -\frac{1}{2}+H)$ (b), (d) (green line), measured with incident neutron energies of $E_{\mathrm{i}}\,{=}\,30.0$ and 17.9 meV, where the scattering intensities were averaged for $-5\leq L\leq7$ to map the scattering intensity in the 2D reciprocal lattice shown in (f), assuming good two-dimensionality. (e) Scattering intensity map in the $(\frac{1}{2}+H, -\frac{1}{2}+H, L)$ plane measured with $E_{\mathrm{i}}\,{=}\,17.9$ meV, where the scattering intensity was averaged for $-0.55\leq K\leq-0.45$ along $\bm{Q}\,{=}\,(-K, K, 0)$. The averaged energy range is $3\leq\hbar\omega\leq10$ meV. The solid lines in (a)--(d) are dispersion curves calculated on the basis of LSWT with $J\,{=}\,12.8$ meV, $J'\,{=}\,-0.043J$, $D^\parallel\,{=}\,0.18J$, and $D^\perp\,{=}\,0.062J$.}
\label{fig:spectra1_CCSF}
\end{figure*} 

\section{Results and Discussion}\label{Results}
\subsection{Cs$_2$Cu$_3$SnF$_{12}$}\label{CCSF}

Figures~\ref{fig:spectra1_CCSF}\,(a)--(d) show excitation spectra of Cs$_2$Cu$_3$SnF$_{12}$ measured along two high-symmetry directions $\bm{Q}\,{=}\,(-\frac{1}{2}-K, -\frac{1}{2}+K)$  and $(\frac{1}{2}+H, -\frac{1}{2}+H)$ at $T\,{=}\,5$ K with incident neutron energies of $E_{\mathrm{i}}\,{=}\,30.0$ and 17.9 meV. We can see four strong excitations around the $\Gamma'$ point, which are ascribed to single-magnon excitations. The excitation energies of these four modes at the $\Gamma'$ point are $\hbar\omega\,{=}\,1.0,\, 9.6,\, 10.7$, and 13.8 meV. The lowest excitation energy $\hbar\omega\,{=}\,1.0$ meV was determined from the scattering data obtained with $E_{\mathrm{i}}\,{=}\,6.3$ meV. The single-magnon excitation arising from the $\Gamma'$ point is clearly confirmed, as observed in a previous study~\cite{Ono2}. 

Figure~\ref{fig:spectra1_CCSF}\,(e) shows the scattering intensity map in the $(\frac{1}{2}+H, -\frac{1}{2}+H, L)$ plane measured with $E_{\mathrm{i}}\,{=}\,17.9$ meV, where the averaged energy range is $3\leq\hbar\omega\leq10$ meV. Strong scattering for $7.5<L<9.5$, which is approximately independent of $H$, is ascribed to phonon excitation. Strong scattering streaks at $H\,{=}\,\frac{1}{2}+n$ with integer $n$ arise mainly from the single-magnon excitations arising from the $\Gamma'$ points. We can confirm from Fig.~\ref{fig:spectra1_CCSF}\,(e) that the single-magnon dispersion curves are independent of $L$; thus, the interlayer exchange interaction is negligible. 

In a previous study~\cite{Ono2}, Ono \textit{et al}. analyzed the dispersion curves of Cs$_2$Cu$_3$SnF$_{12}$ assuming four kinds of the nearest-neighbor exchange interaction on the basis of the $2a\times2a$ enlarged chemical unit cell below $T_{\mathrm{s}}\,{=}\,185$ K. Here, we analyze the dispersion curves assuming a uniform kagome lattice because the nearest-neighbor exchange interactions are deduced to be nearly the same, as mentioned in Section~\ref{Intro}. We use a magnetic model of Cs$_2$Cu$_3$SnF$_{12}$ expressed as
\begin{align}
\mathcal{H}&=\sum_{\langle i,j\rangle} J\left({\bm{S}}_i\cdot{\bm{S}}_j\right) + \sum_{\langle i,j'\rangle} J'\left({\bm{S}}_i\cdot{\bm{S}}_{j'}\right) \nonumber\\ 
&\quad+ \sum_{\langle i,j\rangle} {\bm{D}}_{ij}\cdot\left({\bm{S}}_i\times{\bm{S}}_j\right),
\label{model_CCSF}
\end{align}
where the first and second terms are the nearest- and next-nearest-neighbor exchange interactions, respectively. The third term is the DM interaction between the nearest-neighbor spins. The configurations of the parallel and perpendicular components of the $\bm{D}_{ij}$ vector, $D_{ij}^\parallel$ and $D_{ij}^\perp$, are illustrated in Figs.~\ref{fig:kagome_latt}\,(d) and (e), respectively, which are derived from the crystal structure at room temperature~\cite{Ono}. We calculated the dispersion curves using LSWT. The solid lines in Figs.~\ref{fig:spectra1_CCSF}\,(a)--(d) are dispersion curves calculated using LSWT with $J\,{=}\,12.8$ meV, $J'\,{=}\,-0.043J$, $D^\parallel\,{=}\,0.18J$, and $D^\perp\,{=}\,0.062J$. These exchange parameters are consistent with those obtained by Ono {\it et al.}~\cite{Ono2}. The ferromagnetic exchange interaction $J'$ between the next-nearest-neighbor spins is necessary to describe the dispersion curve of the weakly dispersive branch between 7.5 and 10 meV. The relatively large $D^{\parallel}$ is essential for the description of the weakly dispersive branch. The small $D^\perp$ gives rise to a small gap of ${\Delta}\,{=}\,1.0$\,meV for the dispersive branch arising from the ${\Gamma}^{\prime}$ points and a small split of two branches near 10\,meV at the ${\Gamma}^{\prime}$ points.

\begin{figure*}[t]
\includegraphics[width=17cm, clip]{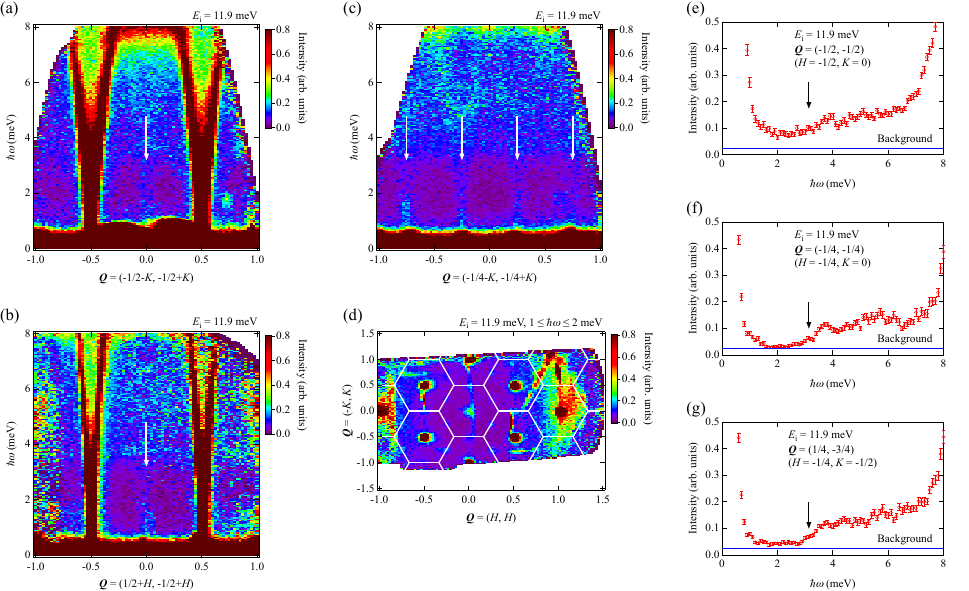}
\caption{(Color online) Excitation spectra of Cs$_2$Cu$_3$SnF$_{12}$ measured at $T\,{=}\,5$ K with incident neutron energy $E_{\mathrm{i}}\,{=}\,11.9$ meV. (a)--(c) Energy--momentum maps of the scattering intensity along $\bm{Q}\,{=}\,(-\frac{1}{2}-K, -\frac{1}{2}+K)$, $(\frac{1}{2}+H, -\frac{1}{2}+H)$ and $(-\frac{1}{4}-K, -\frac{1}{4}+K)$, respectively, where the scattering intensities were averaged for $-5\leq L\leq7$ to map the scattering intensity. Vertical white arrows in (a)--(c) indicate weak excitations arising from the M points. (d) Constant-energy slice of scattering intensity, where the averaged energy range is $1\leq\hbar\omega\leq2$ meV. Solid white lines are boundaries of the elementary BZs for the uniform kagome lattice. (e)--(g) Scattering intensities as a function of energy measured with incident neutron energy $E_{\mathrm{i}}\,{=}\,11.9$ meV for $\bm{Q}\,{=}\,(-\frac{1}{2}, -\frac{1}{2})$, $(-\frac{1}{4}, -\frac{1}{4})$, and $(\frac{1}{4}, -\frac{3}{4})$, which are located at $(H\,{=}\,-\frac{1}{2}, K\,{=}\,0)$, $(H\,{=}\,-\frac{1}{4}, K\,{=}\,0)$, and $(H\,{=}\,-\frac{1}{4}, K\,{=}\,-\frac{1}{2})$, respectively, in the 2D reciprocal lattice shown in Fig.~\ref{fig:spectra1_CCSF}\,(f). Horizontal lines are the background levels estimated from the scattering intensity in the vicinity of the $\Gamma$ points. Vertical arrows in (e)--(g) indicate lower energy cut of the excitation continuum at approximately 3 meV, which is nearly equal to $0.15J$.}
\label{fig:spectra2_CCSF}
\end{figure*} 

Low-energy dispersion curves below 10 meV appear to be well reproduced by the LSWT calculation. From the linear dispersion in the vicinity of the $\Gamma'$ points, the nearest-neighbor exchange constant $J$ can be determined to be $J_{\mathrm{disp}}\,{=}\,12.8$ meV. However, this exchange constant is significantly smaller than $J_{\mathrm{mag}}\,{=}\,20.7$ meV obtained from the analysis of the temperature dependence of magnetic susceptibility using the theoretical result obtained by exact diagonalization for the 24-site kagome cluster~\cite{Ono}. We consider that the true exchange constant $J$ is close to $J_{\mathrm{mag}}$ obtained from the magnetic susceptibility; thus, the quantum renormalization factor is evaluated to be $R\,{=}\,J_{\mathrm{disp}}/J_{\mathrm{mag}}\,{=}\,0.62$. From these results, we can confirm that the excitation energy of the single magnon is largely renormalized downward by the quantum many-body effect, as concluded in Ref.~\cite{Ono2}. This quantum renormalization effect of excitation energy observed in $S\,{=}\,1/2$ KLHAF Cs$_2$Cu$_3$SnF$_{12}$ contrasts with that observed in the $S\,{=}\,1/2$ TLHAF Ba$_3$CoSb$_2$O$_9$, in which the energy of the single-magnon excitation arising from the K point is not renormalized downward in the vicinity of the K point~\cite{Ma,Ito,Macdougal}. Note that no such large downward renormalization of single-magnon excitation energy was observed in an $S\,{=}\,5/2$ KLHAF KFe$_3$(OH)$_6$(SO$_4$)$_2$~\cite{Matan4,Yildirim}. 

For high-energy single-magnon excitations above 10 meV, there is a significant difference between the experimental and LSWT results. Single-magnon excitations near the zone boundary, which are expected at $\hbar\omega\simeq15$ meV from LSWT calculations, appear to significantly broaden out. This suggests the decay of high-energy magnons near the zone boundary, as discussed in Refs.~\cite{Chernyshev2,Chernyshev3}.

The fourth strong single-magnon excitation for $\hbar\omega\simeq14$ meV in the vicinity of the $\Gamma'$ point cannot be explained by LSWT because it derives only three transverse modes. Recently, the magnetic excitations in $S\,{=}\,1/2$ TLHAF was discussed by a fermionic approach of spinon excitations from the ordered ground state with the RVB quantum fluctuation~\cite{Zhang}. Dispersion curves of single-magnon excitations observed in Ba$_3$CoSb$_2$O$_9$ were mostly explained by this theory~\cite{Ito,Macdougal}, in which single-magnon excitations are described as bound states of two spinons. The theory predicts three low-energy transverse modes and one high-energy longitudinal mode. The high-energy longitudinal mode has strong intensity around the K point. This theory will be applicable to the magnetic excitations in Cs$_2$Cu$_3$SnF$_{12}$. The K point in the case of the triangular-lattice antiferromagnet corresponds to the $\Gamma'$ point for Cs$_2$Cu$_3$SnF$_{12}$ with the $\bm{q}\,{=}\,0$ ordered ground state. Thus, we deduce that the strong single-magnon excitation for $\hbar\omega\,{\simeq}\,14$ meV in the vicinity of the $\Gamma'$ point is the amplitude mode and that its strong intensity is caused by that the single-magnon excitations are composed of bound spinons. To confirm the amplitude mode, the measurement of the mode polarization is necessary. 

Figures~\ref{fig:spectra2_CCSF}\,(a)--(c) show the excitation spectra of Cs$_2$Cu$_3$SnF$_{12}$ along $\bm{Q}\,{=}\,(-\frac{1}{2}-K, -\frac{1}{2}+K)$, $(\frac{1}{2}+H, -\frac{1}{2}+H)$ and $(-\frac{1}{4}-K, -\frac{1}{4}+K)$, respectively, measured at $T\,{=}\,5$ K with $E_{\mathrm{i}}\,{=}\,11.9$ meV. Besides the strong single-magnon excitation arising from the $\Gamma'$ point, weak excitations arising from the M points are also observed, as indicated by vertical white arrows. Figure~\ref{fig:spectra2_CCSF}\,(d) shows a constant-energy slice of scattering intensity, where the averaged energy range is $1\leq\hbar\omega\leq2$ meV. Bright and dark spots are observed at the $\Gamma'$ and M points, respectively. Thus, it is confirmed that the strong and weak excitations rise from the $\Gamma'$ and M points, respectively. 

There are two possible origins of the weak excitations arising from the M points. One is the ghost modes of the main single-magnon modes caused by the enlargement of the chemical unit cell below the structural phase transition temperature $T_{\mathrm{t}}\,{=}\,185$ K. The crystal lattice is actually $a\,{\times}\,2a$ enlarged in the $ab$ plane below $T_{\mathrm{t}}$~\cite{Downie2,Matan3}. Because the lattice distortion is small and there are three kinds of equivalent structural domain, the diffraction spots for the crystal lattice appear to be those of the $2a\,{\times}\,2a$ enlarged unit cell, as assumed in a previous study~\cite{Ono2}. In such a case, the M points of elementary BZs for a uniform kagome lattice become the zone centers in contracted BZs. Consequently, weak ghost modes of strong single-magnon excitation modes can occur at the M points, as observed in an analogous compound, Rb$_2$Cu$_3$SnF$_{12}$~\cite{Matan2}.

\begin{figure*}[t]
\includegraphics[width=17.5cm, clip]{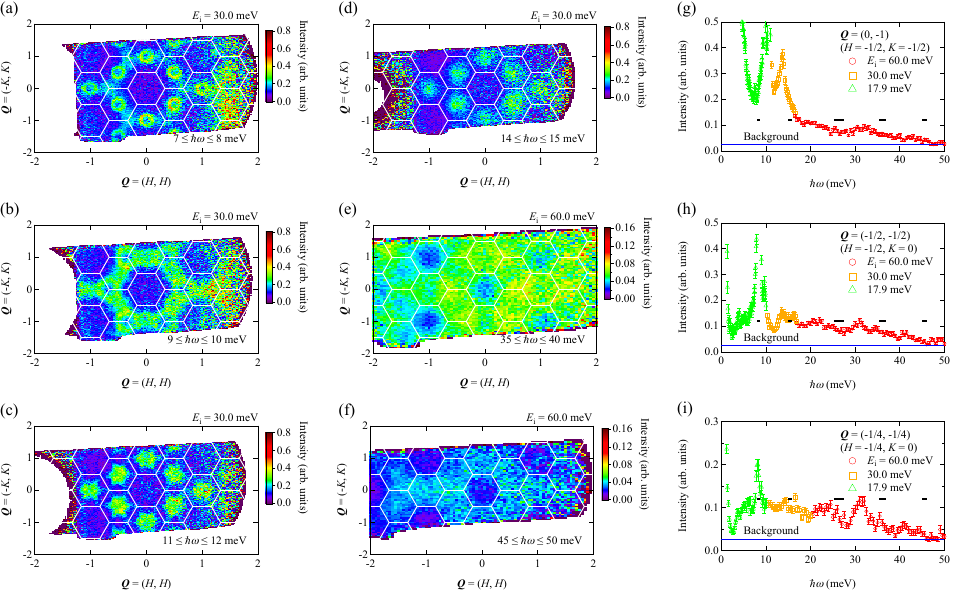}
\caption{(Color online) (a)--(f) Constant-energy slices of scattering intensity of Cs$_2$Cu$_3$SnF$_{12}$ measured at $T\,{=}\,5$ K. (a)--(d), and (e) and (f) are the results obtained using the scattering data measured with incident neutron energies of $E_{\mathrm{i}}\,{=}\,30.0$ and 60.0 meV, respectively, where the scattering intensities were averaged in momentum ranges of $-5\leq L\leq7$ and $-10\leq L\leq14$ for $E_{\mathrm{i}}\,{=}\,30.0$ and 60.0 meV, respectively. Averaged energy ranges are shown in the figures. Solid white lines are BZ boundaries. (g)--(i) Scattering intensities as a function of energy measured for $\bm{Q}\,{=}\,(0, -1)$, $(-\frac{1}{2}, -\frac{1}{2})$, and $(-\frac{1}{4}, -\frac{1}{4})$, which are located at $(H\,{=}\,-\frac{1}{2}, K\,{=}\,-\frac{1}{2})$, $(H\,{=}\,-\frac{1}{2}, K\,{=}\,0)$, and $(H\,{=}\,-\frac{1}{4}, K\,{=}\,0)$, respectively, in the 2D reciprocal lattice space shown in Fig.~\ref{fig:spectra1_CCSF}\,(f). Here, scattering data obtained with incident neutron energy $E_{\mathrm{i}}\,{=}\,17.9$, 30.0, and 60.0 meV are combined, so that the low-energy structures of the excitation spectra are clearly visible. Horizontal lines and bars are the background levels and energy resolution, respectively. $\bm{Q}$-independent peaks at 31 meV in (g)--(i) are nonmagnetic peaks due to the phonons in Cs$_2$Cu$_3$SnF$_{12}$.}
\label{fig:slice_30_60_CCSF}
\end{figure*} 

The other possible origin is the excitations from a component of the spin liquid state such as the gapless $U(1)$ Dirac spin liquid and the gapped $Z_2$ spin liquid, which remains in the ground state as quantum fluctuations. Recent theories based on the spinon excitations from the spin liquid ground state demonstrated that the strong and weak magnetic excitations rise from the $\Gamma'$ and M points in the extended BZs shown in Fig.~\ref{fig:kagome_latt}\,(a)~\cite{Punk,Zhang2,Ferrari2}. Although Cs$_2$Cu$_3$SnF$_{12}$ has an ordered ground state, it seems possible that the spin liquid state remains as quantum fluctuations as in the case of the $S\,{=}\,1/2$ TLHAF~\cite{Ferrari,Zhang}. Although the possibility of the ghost mode due to the enlargement of the chemical unit cell appears to be higher because the ghost mode was actually observed in an analogous compound Rb$_2$Cu$_3$SnF$_{12}$~\cite{Matan2}, the second possibility remains. To examine which possibility is correct, INS experiments using a model system without lattice distortion are necessary.

As seen from Figs.~\ref{fig:spectra2_CCSF}\,(a)--(c), the scattering intensity changes at approximately 3 meV irrespective of the momentum transfer $\bm{Q}$ except in the vicinities of the $\Gamma'$ and M points. Figures~\ref{fig:spectra2_CCSF}\,(e)--(g) show the scattering intensities as a function of energy measured with $E_{\mathrm{i}}\,{=}\,11.9$ meV for $\bm{Q}\,{=}\,(-\frac{1}{2}, -\frac{1}{2})$, $(-\frac{1}{4}, -\frac{1}{4})$, and $(\frac{1}{4}, -\frac{3}{4})$. The steep increase in scattering intensity below 1 meV is due to the tail of incoherent scattering. The scattering intensity, which is almost at the background level in the energy range of $1.5\leq\hbar\omega\leq2.5$ meV except in the vicinity of the $\Gamma'$ and M points, increases rapidly at approximately 3 meV and increases gradually with increasing excitation energy, which is a characteristic of the excitation continuum. As seen from Figs.~\ref{fig:spectra2_CCSF}\,(a)--(c), the scattering intensity is almost independent of $\bm{Q}$ in a wide area of the BZ centered at the $\Gamma'$ point except in the vicinity of the $\Gamma'$ point. These results indicate that there is an almost structureless excitation continuum with a lower bound of $0.15J\simeq3$ meV. 

Figures~\ref{fig:slice_30_60_CCSF}\,(a)--(f) show constant-energy slices of scattering intensity in Cs$_2$Cu$_3$SnF$_{12}$ plotted in the 2D reciprocal lattice space. Scattering intensities in the BZs centered at the $\Gamma$ points are absent owing to the kagome geometry. For the dynamical structure factor $\mathcal{S}(\bm{Q}, \omega)$ of the kagome-lattice magnet, there is the relationship of $\mathcal{S}(\bm{Q}, \omega)\,{=}\,\mathcal{S}(\bm{Q}+2\bm{G}, \omega)$, where $\bm{G}$ is the reciprocal lattice vector. The weak scattering intensities in the BZs centered at $(H\,{=}\,1, K\,{=}\,\pm1)$ are of phonon origin. 

The intensity distribution below 12 meV is mainly attributed to the single-magnon excitations. Strong scattering in this energy range has clear structures in the 2D reciprocal lattice space. The structures are circular, X-like, and six-pointed star patterns centered at the $\Gamma'$ points for $7\,{\leq}\,\hbar\omega\,{\leq}\,8$ meV, $9\,{\leq}\,\hbar\omega\,{\leq}\,10$ meV, and $11\,{\leq}\,\hbar\omega\,{\leq}\,12$ meV, respectively. These intensity maps appear to be consistent with those calculated by spin wave theory for a large spin KLHAF with the DM interaction~\cite{Chernyshev3}. In the energy range of the excitation continuum above 14 meV, the scattering intensity does not exhibit a marked structure in the BZs centered at the $\Gamma'$ points. This almost structureless high-energy excitation in Cs$_2$Cu$_3$SnF$_{12}$ is similar to that observed in herbertsmithite ZnCu$_3$(OH)$_6$Cl$_2$~\cite{Han,Han3}.

Figures~\ref{fig:slice_30_60_CCSF}\,(g)--(i) show scattering intensities as a function of energy measured for $\bm{Q}\,{=}\,(0, -1)$, $(-\frac{1}{2}, -\frac{1}{2})$, and $(-\frac{1}{4}, -\frac{1}{4})$. $\bm{Q}$-independent peaks at 31 meV in these figures are nonmagnetic peaks due to the phonons in Cs$_2$Cu$_3$SnF$_{12}$. Excitation peaks at $\hbar\omega\,{=}\,8\,{-}\,12$ meV are intrinsic peaks due to the single-magnon excitations. The steep increase in scattering intensity below 7 meV for $\bm{Q}\,{=}\,(0, -1)$ is caused by the tail of the magnetic Bragg peak and strong low-energy single-magnon excitation. The steep increase in scattering intensities below 2 meV for $\bm{Q}\,{=}\,(-\frac{1}{2}, -\frac{1}{2})$ and $(-\frac{1}{4}, -\frac{1}{4})$ is due to the tail of incoherent scattering. The sharp peaks at $\hbar\omega\,{\simeq}\,10$ and 13 meV for $\bm{Q}\,{=}\,(0, -1)$, and at $\hbar\omega\,{\simeq}\,8$ meV for $\bm{Q}\,{=}\,(-\frac{1}{2}, -\frac{1}{2})$ and $(-\frac{1}{4}, -\frac{1}{4})$ are single-magnon excitation peaks.

\begin{figure*}[t]
\includegraphics[width=17.5cm, clip]{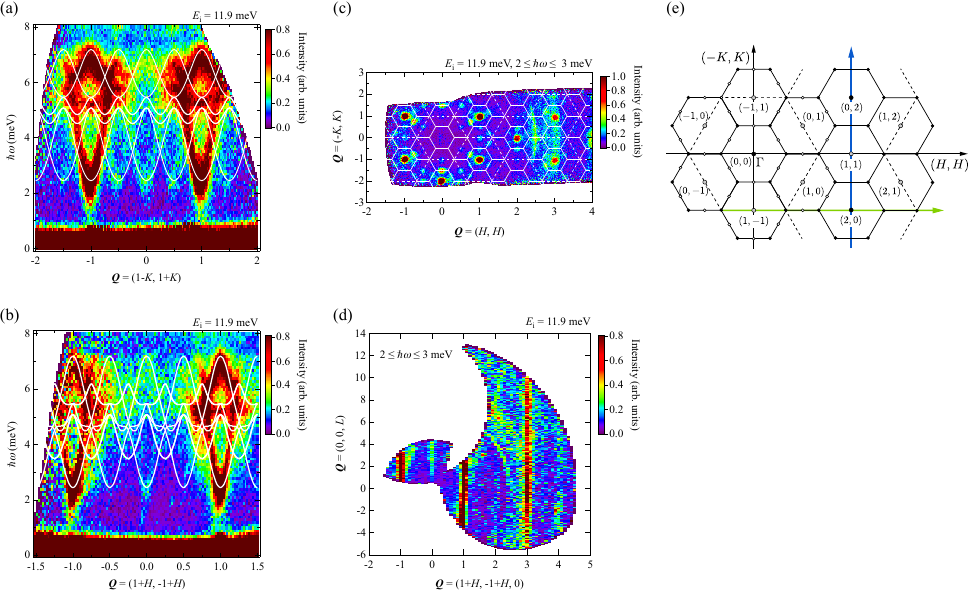}
\caption{(Color online) (a) and (b) Energy--momentum maps of the scattering intensity in Rb$_2$Cu$_3$SnF$_{12}$ measured at $T\,{=}\,5$ K with incident neutron energy $E_{\mathrm{i}}\,{=}\,11.9$ meV along two high-symmetry directions $\bm{Q}\,{=}\,(1-K, 1+K)$ and $(1+H, -1+H)$ indicated by blue and green arrows in the 2D reciprocal lattice shown in (e), respectively, where the solid and dashed lines are elementary BZ boundaries for the $2a\times2a$ enlarged chemical unit cell at room temperature of Rb$_2$Cu$_3$SnF$_{12}$ and the uniform kagome lattice, respectively. Here, the scattering intensities were averaged for $-3\,{\leq}\,L\,{\leq}\,5$ assuming good two-dimensionality. The solid lines are dispersion curves for main singlet--triplet excitations and their ghost modes calculated by the method described in Ref.~\cite{Matan2} with exchange parameters shown in the text. (c) Constant-energy slices of scattering intensity with the averaged energy range of $2\,{\leq}\,\hbar\omega\,{\leq}\,3$ meV, which shows the center positions of strong singlet--triplet excitations and their ghost modes. The solid and dashed white lines are elementary BZ boundaries for the $2a\times2a$ enlarged chemical unit cell at room temperature and the uniform kagome lattice, respectively. (d) Scattering intensity map in the $(1+H, -1+H, L)$ plane measured with $E_{\mathrm{i}}\,{=}\,11.9$ meV, where the scattering intensity was averaged for $-1.05\,{\leq}\,K\,{\leq}\,-0.95$ along $\bm{Q}\,{=}\,(-K, K, 0)$ and for the energy range of $2\,{\leq}\,\hbar\omega\,{\leq}\,3$ meV.}
\label{fig:spectra1_RCSF}
\end{figure*} 

As can be seen from Figs.~\ref{fig:slice_30_60_CCSF}\,(g)--(i), the excitation continuum extends up to approximately 50 meV, which is nearly equal to $2.5J$. Because the lower bound of the excitation continuum is $0.15J\,{\simeq}\,3$ meV, we can say that the broad excitation continuum without a marked structure spreads in a wide energy range from $0.15J$ to approximately $2.5J$. We can see from Figs.~\ref{fig:slice_30_60_CCSF}\,(g)--(i), that the spectral weight of the excitation continuum is sufficiently larger than that of the single-magnon excitations, as observed in the $S\,{=}\,1/2$ triangular-lattice Heisenberg-like antiferromagnet Ba$_3$CoSb$_2$O$_9$~\cite{Ito,Macdougal}. Note that recent theories based on the fermionic approach of spinon excitations from the spin liquid ground state demonstrated that a broad and almost featureless excitation continuum extends to an energy as high as $2.7J$~\cite{Zhang2,Ferrari2}. This theoretical upper bound energy of the excitation continuum is consistent with $2.5J$ observed in Cs$_2$Cu$_3$SnF$_{12}$. From the experimental results for Cs$_2$Cu$_3$SnF$_{12}$ as well as recent theoretical results~\cite{Zhang2,Ferrari2}, we can deduce that the intense excitation continuum with a wide energy range observed in Cs$_2$Cu$_3$SnF$_{12}$ originates from the spinon excitations and that the spin liquid state remains in the ordered ground state as the quantum fluctuation. It is also notable that the inside area between two single-magnon branches arising from the $\Gamma'$ point is filled with a strong excitation continuum, as observed in Figs.~\ref{fig:spectra1_CCSF}\,(a)--(d) and Figs.~\ref{fig:spectra2_CCSF}\,(a) and (b).

\subsection{Rb$_2$Cu$_3$SnF$_{12}$}

In this subsection, we analyze the excitation data for Rb$_2$Cu$_3$SnF$_{12}$ on the basis of the $2a\times2a$ enlarged chemical unit cell at room temperature~\cite{Morita}. The solid and dashed lines in Fig.~\ref{fig:spectra1_RCSF}\,(e) are elementary BZ boundaries for the $2a\times2a$ enlarged chemical unit cell and the uniform kagome lattice, respectively. Figures~\ref{fig:spectra1_RCSF}\,(a) and (b) show excitation spectra in Rb$_2$Cu$_3$SnF$_{12}$ measured at $T\,{=}\,5$ K with the incident neutron energy of $E_{\mathrm{i}}\,{=}\,11.9$ meV along two high-symmetry directions indicated by blue and green arrows in the 2D reciprocal lattice shown in Fig.~\ref{fig:spectra1_RCSF}\,(e). In Figs.~\ref{fig:spectra1_RCSF}\,(a) and (b), we can see strong singlet--triplet excitations centered at $\bm{Q}\,{=}\,(2, 0)$, $(0, 2)$, and $(0, -2)$, as observed in previous studies~\cite{Matan, Matan2}. We can also clearly see the ghost modes of strong singlet--triplet excitations centered at $\bm{Q}\,{=}\,(\frac{3}{2}, \frac{1}{2})$, $(1, 1)$, $(\frac{1}{2}, \frac{3}{2})$, and $(1, {-}1)$, which originate from the contraction of the BZs caused by the enlargement of the chemical unit cell at room temperature and below the structural phase transition at $T_{\mathrm{t}}\,{=}\,215$ K~\cite{Matan,Downie}. 

\begin{figure*}[t]
\includegraphics[width=17.5cm, clip]{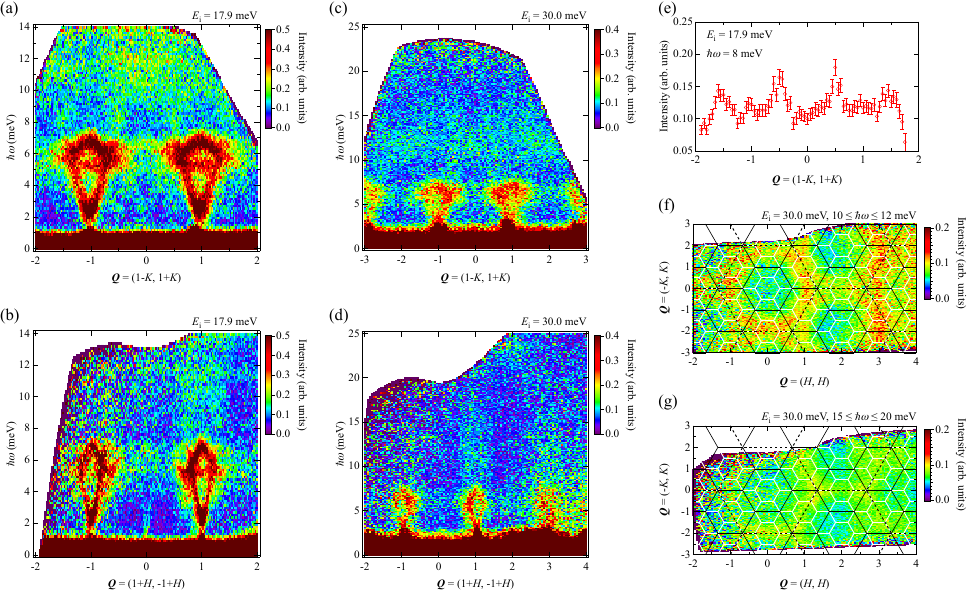}
\caption{(Color online) Energy--momentum maps of the scattering intensity in Rb$_2$Cu$_3$SnF$_{12}$ measured at $T\,{=}\,5$ K with incident neutron energy $E_{\mathrm{i}}\,{=}\,17.9$ meV ((a) and (b)) and 30.0 meV ((c) and (d)) along two high-symmetry directions $\bm{Q}\,{=}\,(1-K, 1+K)$ and $(1+H, -1+H)$ indicated by blue and green arrows in the 2D reciprocal lattice shown in Fig.~\ref{fig:spectra1_RCSF}\,(e), respectively, where the scattering intensities were averaged over $-3\,{\leq}\,L,{\leq}\,5$ for 17.9 meV and $-5\,{\leq}\,L\,{\leq}\,5$ for 30.0 meV. (e) Scattering intensity measured at $\hbar\omega\,{=}\,8.0$ meV along $\bm{Q}\,{=}\,(1-K, 1+K)$ with incident neutron energy $E_{\mathrm{i}}\,{=}\,17.9$ meV, where the scattering intensity was averaged for $7.8\,{\leq}\,\hbar\omega\,{\leq}\,8.2$ meV.
(f) and (g) Constant-energy slices of scattering intensity of Rb$_2$Cu$_3$SnF$_{12}$ measured at $T\,{=}\,5$ K with incident neutron energy $E_{\mathrm{i}}\,{=}\,30.0$ meV. Averaged energy ranges are shown in the figures. The white solid lines are elementary BZ boundaries for the $2a\times2a$ enlarged chemical unit cell at room temperature of Rb$_2$Cu$_3$SnF$_{12}$, and the black solid and dashed lines are elementary and extended BZ boundaries for the uniform kagome lattice, respectively.}
\label{fig:spectra2_RCSF}
\end{figure*} 

\begin{figure*}[t]
\includegraphics[width=17.5cm, clip]{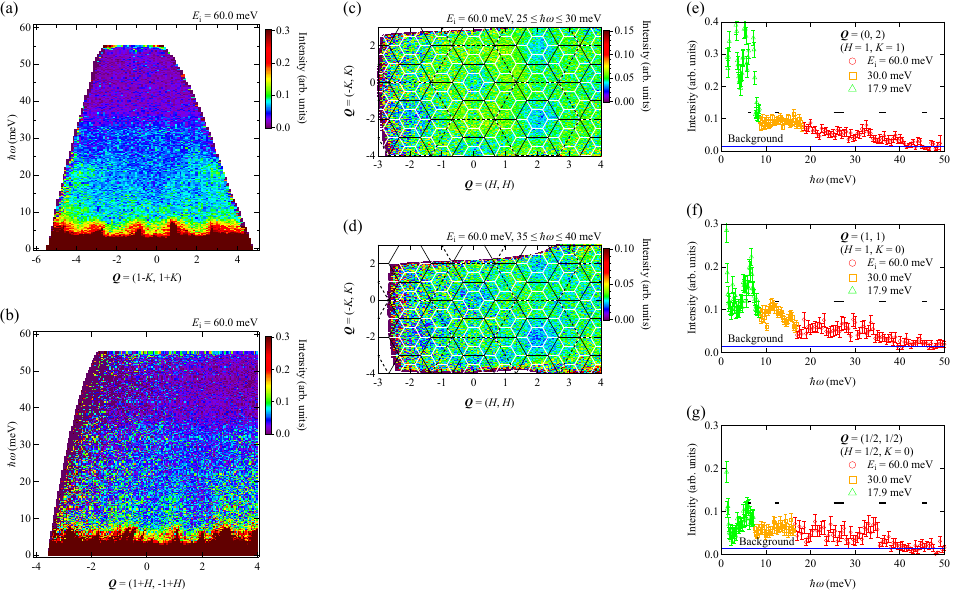}
\caption{(Color online) (a) and (b) Energy--momentum maps of the scattering intensity in Rb$_2$Cu$_3$SnF$_{12}$ measured at $T\,{=}\,5$ K with incident neutron energy $E_{\mathrm{i}}\,{=}\,60.0$ meV along two high-symmetry directions $\bm{Q}\,{=}\,(1-K, 1+K)$ and $(1+H, -1+H)$ indicated by blue and green arrows in the 2D reciprocal lattice shown in Fig.~\ref{fig:spectra1_RCSF}\,(e), respectively, where the scattering intensities were averaged for $-5\,{\leq}\,L\,{\leq}\,15$. $\bm{Q}$-independent scattering at $\hbar\omega\simeq32$ meV are nonmagnetic peaks due to the phonons in Rb$_2$Cu$_3$SnF$_{12}$.
(c) and (d) Constant-energy slices of scattering intensity of Rb$_2$Cu$_3$SnF$_{12}$ measured at $T\,{=}\,5$ K with incident neutron energy $E_{\mathrm{i}}\,{=}\,60.0$ meV. Averaged energy ranges are shown in the figures. The white solid lines are elementary BZ boundaries for the $2a\times2a$ enlarged chemical unit cell at room temperature for Rb$_2$Cu$_3$SnF$_{12}$, and the black solid and dashed lines are elementary and extended BZ boundaries for the uniform kagome lattice. (e)--(g) Scattering intensities as a function of energy measured for $\bm{Q}\,{=}\,(0, 2)$, $(1,1)$, and $(\frac{1}{2}, \frac{1}{2})$, which are located at $(H\,{=}\,1, K\,{=}\,1)$, $(H\,{=}\,1, K\,{=}\,0)$, and $(H\,{=}\,\frac{1}{2}, K\,{=}\,0)$, respectively, in the 2D reciprocal lattice space shown in Fig.~\ref{fig:spectra1_RCSF}\,(e). Here, scattering data obtained with incident neutron energy $E_{\mathrm{i}}\,{=}\,17.9$, 30.0, and 60.0 meV are combined, so that the low-energy structures of the excitation spectra are clearly visible. Horizontal lines and bars are the background levels and energy resolution, respectively. $\bm{Q}$-independent peaks at $\hbar\omega\,{\simeq}\,32$ meV are nonmagnetic peaks due to the phonons in Rb$_2$Cu$_3$SnF$_{12}$.}
\label{fig:spectra3_RCSF}
\end{figure*} 

For the strong singlet--triplet excitation, the lowest-energy excitation occurs at the $\Gamma$ point in the 2D reciprocal lattice for the $2a\times2a$ enlarged unit cell, although the lowest-energy excitation is expected to occur at the K point within the Heisenberg model~\cite{Matan,Yang2}. This is because the out-of-plane component $D_{ij}^\parallel$ of the DM interaction splits the triply degenerate triplet excitations into two levels, $S^z\,{=}\,0$ and $S^z\,{=}\,{\pm}1$ branches, and the energy of the $S^z\,{=}\,{\pm}1$ branch becomes minimum at the $\Gamma$ point with increasing magnitude of $D_{ij}^\parallel$~\cite{Matan,Hwang2}.

Figure~\ref{fig:spectra1_RCSF}\,(c) shows constant-energy slices of scattering intensity with the averaged energy range of $2\,{\leq}\,\hbar\omega\,{\leq}\,3$ meV. The center positions of strong singlet--triplet excitations and their ghost modes are clearly observed in Fig.~\ref{fig:spectra1_RCSF}\,(c). Strong singlet--triplet excitations are centered at $\bm{Q}\,{=}\,(4m+2, 2n)$ or $(2m, 4n+2)$ with the integers $m$ and $n$, which correspond to the $\Gamma'$ points for the extended BZ for the uniform kagome lattice shown in Fig.~\ref{fig:kagome_latt}\,(a). Ghost modes are centered at the zone centers and the middle points (M point) of neighboring zone centers of the 2D reciprocal lattice for the $2a\times2a$ enlarged chemical unit cell at room temperature in Rb$_2$Cu$_3$SnF$_{12}$. The ghost modes centered at the zone centers can be produced by the contraction of the BZs caused by the $2a\times2a$ enlargement of the chemical unit cell at room temperature. On the other hand, the ghost modes centered at the M points originate from the further enlargement of the chemical unit cell below $T_{\mathrm{t}}\,{=}\,215$ K~\cite{Matan,Downie}. The crystal structure below $T_{\mathrm{t}}$ has been considered to be a triclinic structure ($P{\bar{1}}$)~\cite{Downie}. However, the structure is very close to a trigonal structure with the cell dimensions $4a\times4a$ as compared with the uniform kagome lattice~\cite{Matan,Downie}. In general, there are three kinds of equivalent structural domain; thus, the diffraction spots for the crystal lattice below $T_{\mathrm{t}}$ appear to be those for the $4a\times4a$ enlarged unit cell, as assumed in a previous study~\cite{Matan2}. Because the M points become the zone centers in the 2D reciprocal lattice for the $4a\times4a$ enlarged chemical unit cell, ghost modes centered at the M points can be observed. As seen from Fig.~\ref{fig:spectra1_RCSF}\,(c), the strongest ghost modes are centered at $\bm{Q}\,{=}\,(\frac{3}{2}, 0)$ $(H\,{=}\,\frac{3}{4}, K\,{=}\,-\frac{3}{4})$ and its equivalent positions. In the previous study~\cite{Matan2}, only the strongest ghost modes were observed.

The model Hamiltonian for the $2a\times2a$ enlarged chemical unit cell at room temperature, which contains twelve spins, is expressed as 
\begin{equation}
\mathcal{H}=\sum_{\langle i,j\rangle} J_{ij}\left(\bm{S}_i\cdot\bm{S}_j\right) + \sum_{\langle i,j\rangle} \bm{D}_{ij}\cdot\left(\bm{S}_i\times\bm{S}_j\right),
\label{model_RCSF}
\end{equation} 
where the configuration of the nearest-neighbor exchange interactions and the DM interactions are shown in Figs.~\ref{fig:kagome_latt}\,(c)--(e) and their values were determined to be $J_1\,{=}\,18.6$ meV, $J_2\,{=}\,0.95J_1$, $J_3\,{=}\,0.85J_1$, $J_4\,{=}\,0.55J_1$, and $D_\alpha^\parallel\,{=}\,0.18J_\alpha$~\cite{Matan}. Here, the in-plane component $D_\alpha^\perp$ of the $\bm{D}$ vectors is neglected. For the $4a\times4a$ enlarged chemical unit cell, which can approximate the crystal structure below $T_{\rm t}\,{\,{=}\,}\,215$ K~\cite{Matan,Downie}, the magnitudes of $J_{ij}$ and $D_\alpha^\parallel$ are slightly modified. This perturbation gives rise to the ghost modes. The solid lines in Figs.~\ref{fig:spectra1_RCSF}\,(a) and (b) are the dispersion curves for the singlet--triplet excitations and their ghost modes, which are the same as those presented in Ref.~\cite{Matan2}.

Figure~\ref{fig:spectra1_RCSF}\,(d) shows the scattering intensity map in the $(1\,{+}\,H, {-}1\,{+}\,H, L)$ plane for the energy range of $2\leq\hbar\omega\leq3$ meV. Strong scattering streaks at $H\,{=}\,2n+1$ with the integer $n$ arise from the singlet--triplet excitations. From Fig.~\ref{fig:spectra1_RCSF}\,(d), we can confirm that the dispersion curves for the singlet--triplet excitations are independent of $L$; thus, the interlayer exchange interaction is negligible. The weak curved scattering observed between $H\,{=}\,1$ and 3 is considered to be spurious scattering from sample environment.

Figures~\ref{fig:spectra2_RCSF}\,(a)--(d) show energy--momentum maps of the scattering intensity in Rb$_2$Cu$_3$SnF$_{12}$ measured at $T\,{\,{=}\,}\,5$\,K with $E_{\mathrm{i}}\,{=}\,17.9$ meV ((a) and (b)) and 30 meV ((c) and (d)) along two high-symmetry directions indicated by blue and green arrows in the 2D reciprocal lattice shown in Fig.~\ref{fig:spectra1_RCSF}\,(e), respectively. In the energy range of $7\,{<}\,\hbar\omega\,{<}\,15$ meV, which is just above the energy range of the singlet--triplet excitation ($2\,{<}\,\hbar\omega\,{<}\,7$ meV), we can see the structured excitation continuum. Figure~\ref{fig:spectra2_RCSF}\,(e) shows the scattering intensity measured at $\hbar\omega\,{=}\,8.0$ meV along $\bm{Q}\,{=}\,(1-K, 1+K)$ with incident neutron energy $E_{\mathrm{i}}\,{=}\,17.9$ meV. Strong scattering occurs at $K\,{=}\,n+\frac{1}{2}$ with the integer $n$. As can be seen from Figs.~\ref{fig:spectra2_RCSF}\,(a) and (c), the scattering intensity has a broad maximum at $\hbar\omega\,{\simeq}\,12$ meV, which is approximately twice the energy of the upper branch of the singlet--triplet excitation. Thus, the structured excitation continuum ranging from 7 meV to 15 meV appears to be two-triplet excitations from the pinwheel VBS ground state.

Figures~\ref{fig:spectra2_RCSF}\,(f) and (g) show constant-energy slices of the scattering intensity of Rb$_2$Cu$_3$SnF$_{12}$ measured at $T\,{=}\,5$ K with $E_{\mathrm{i}}\,{=}\,30.0$ meV in the energy ranges of $10\,{\leq}\,\hbar\omega\,{\leq}\,12$ and $15\,{\leq}\,\hbar\omega\,{\leq}\,20$ meV, respectively. In these energy ranges of the excitation continuum, strong scattering occurs in the vicinity of the extended BZ boundaries for the uniform kagome lattice. 

Figures~\ref{fig:spectra3_RCSF}\,(a) and (b) show energy--momentum maps of the scattering intensity in Rb$_2$Cu$_3$SnF$_{12}$ measured at $T\,{=}\,5$ K with $E_{\mathrm{i}}\,{=}\,60.0$ meV along two high-symmetry directions indicated by blue and green arrows in the 2D reciprocal lattice shown in Fig.~\ref{fig:spectra1_RCSF}\,(e), respectively. We can see that the excitation continuum without a marked structure extends at least up to 35 meV.

Figures~\ref{fig:spectra3_RCSF}\,(c) and (d) show constant-energy slices of scattering intensity of Rb$_2$Cu$_3$SnF$_{12}$ measured at $T\,{=}\,5$ K with incident neutron energy $E_{\mathrm{i}}\,{=}\,60.0$ meV. In the energy range of $25\,{\leq}\,\hbar\omega\,{\leq}\,30$ meV, the scattering intensity of the excitation continuum distributes all over the BZs centered at $\Gamma'$ points in the 2D reciprocal lattice space for the uniform kagome lattice and does not exhibits a marked structure. We can see that for $35\,{\leq}\,\hbar\omega\,{\leq}\,40$ meV, the scattering intensities in the BZs centered at $\bm{Q}\,{=}\,(4m, 4n)$ with the integers $m$ and $n$, which correspond to $\Gamma$ points for the extended BZs in the 2D reciprocal lattice for the uniform kagome lattice, are weaker than those in the other BZs. This indicates that the excitation continuum extends up to approximately 40 meV ($\,{=}\,2.56J_{\mathrm{avg}}$). The broad high-energy excitation spectrum without a marked structure in Rb$_2$Cu$_3$SnF$_{12}$ is similar to the high-energy excitation spectrum observed in Cs$_2$Cu$_3$SnF$_{12}$ and the excitation spectrum observed in herbertsmithite ZnCu$_3$(OH)$_6$Cl$_2$~\cite{Han,Han3}.

Figures~\ref{fig:spectra3_RCSF}\,(e)--(g) show scattering intensities as a function of energy measured for $\bm{Q}\,{=}\,(0, 2)$, $(1,1)$, and $(\frac{1}{2}, \frac{1}{2})$. Sharp excitation peaks below 8 meV arise from the singlet--triplet excitations from the pinwheel VBS state and their ghost modes. We can confirm from Figs.~\ref{fig:spectra3_RCSF}\,(e)--(g) that the upper bound energy of the excitation continuum is approximately 40 meV, which is 2.56 times larger than the average of the nearest-neighbor exchange interactions $J_{\mathrm{avg}}\,{=}\,15.6$ meV. This upper bound energy of the excitation continuum observed in Rb$_2$Cu$_3$SnF$_{12}$ is almost the same as that observed in Cs$_2$Cu$_3$SnF$_{12}$ and consistent with the theoretical upper bound energy of $2.7J$ obtained by a fermionic approach of spinon excitations from the spin liquid ground state~\cite{Zhang2,Ferrari2}. The spectrum of excitation energy for $\bm{Q}\,{=}\,(\frac{1}{2}, \frac{1}{2})$, in which the weight of the singlet--triplet excitation is much smaller than the weight of the excitation continuum, is similar to that for $\bm{Q}\,{=}\,(-\frac{1}{4}, -\frac{1}{4})$ in Cs$_2$Cu$_3$SnF$_{12}$ shown in Fig.~\ref{fig:slice_30_60_CCSF}\,(i). These observations indicate that the high-energy excitation in $S\,{=}\,1/2$ Heisenberg-like kagome-lattice antiferromagnets has common characteristics, irrespective of the ground state.

\section{Conclusions}
We have presented the results of INS experiments on the $S\,{=}\,1/2$ KLHAFs Cs$_2$Cu$_3$SnF$_{12}$ with the $\bm{q}\,{=}\,0$ ordered ground state attributable to the DM interaction and Rb$_2$Cu$_3$SnF$_{12}$ with the gapped pinwheel VBS state attributable to the $2a\times2a$ enlarged chemical unit cell.
For Cs$_2$Cu$_3$SnF$_{12}$, we observed four single-magnon excitation modes around the $\Gamma'$ point in the extended BZ. Low-energy three modes are assigned to be transverse modes, whereas the high-energy fourth mode is suggested to be a longitudinal mode. We confirmed that the energy of single-magnon excitations arising from the $\Gamma'$ point is largely renormalized downwards and that the high-energy single-magnon excitations near the zone boundary significantly broaden out. We found that the broad excitation continuum without a marked structure spreads in a wide energy range from $0.15J$ to approximately $2.5J$. In addition, we found weak excitations arising from the M points. Two possibilities are considered as the origins of the weak excitations. One is the ghost modes of the main single-magnon modes caused by the enlargement of the chemical unit cell below a structural phase transition temperature $T_{\mathrm{t}}\,{=}\,185$ K. The other is the spinon excitations from the spin liquid-like component remaining in the ground state. For Rb$_2$Cu$_3$SnF$_{12}$, we confirmed singlet--triplet excitations from the pinwheel VBS state and their ghost modes caused by the enlargement of the chemical unit cell. We found that the excitation continuum in the low-energy range up to approximately $J_{\mathrm{avg}}\,{=}\,15.6$ meV has clear structure, whereas the high-energy excitation continuum above $J_{\mathrm{avg}}$ is almost structureless and extends to approximately $2.6J_{\mathrm{avg}}$. The characteristics of the high-energy excitation continuum are common to both Cs$_2$Cu$_3$SnF$_{12}$ and Rb$_2$Cu$_3$SnF$_{12}$, irrespective of their ground states; thus, these characteristics of the high-energy excitation continuum are considered to be universal in the $S\,{=}\,1/2$ KLHAF. Because these observations are consistent with the theories based on a fermionic approach of spinon excitations from the spin liquid ground state such as the gapless $U(1)$ Dirac spin liquid state and the gapped $Z_2$ spin liquid state~\cite{Zhang2,Ferrari2}, we can deduce that the spin liquid component remains in the ground state as quantum fluctuations in Cs$_2$Cu$_3$SnF$_{12}$ and Rb$_2$Cu$_3$SnF$_{12}$.


\begin{acknowledgments}
We would like to thank K. Matan, T. J. Sato, and K. Nakajima for fruitful discussions. 
This work was supported by Grants-in-Aid for Scientific Research (A) (No.~17H01142) and (C) (No.~19K03711) from the Japan Society for the Promotion of Science (JSPS). The experiments on 4SEASONS were performed with the approval of J-PARC (Proposal Nos. 2018B0099 and 2019B0106).

\end{acknowledgments}




\begin{thebibliography}{999}
\bibitem{herbertsmithite} Cu(OH)$_6$ octahedra in the Zn$^{2+}$ layer are distorted due to the Jahn--Teller effect, which leads to the disorder of OH$^-$ ions that mediate superexchange interaction between Cu$^{2+}$ ions in the kagome layer. Because the bonding angle Cu$^{2+}$$-$\,O\,$-$\,Cu$^{2+}$ is neither 180$^{\circ}$ nor 90$^{\circ}$, but close to 120$^{\circ}$, a change in the bonding angle of merely 10$^{\circ}$ will cause a large change in the exchange constant as observed in $S\,{\,{=}\,}\,1/2$ distorted KLHAF Rb$_2$Cu$_3$SnF$_{12}$~\cite{Morita,Matan,Matan2,Downie}. Thus, the disorder of OH$^-$ ions will give rise to the significant exchange randomness.
\end{thebibliography}
\end{document}